\documentclass[prb,twocolumn,showpacs,preprintnumbers,superscriptaddress]{revtex4-1}
\usepackage[usenames,dvipsnames]{color}
\usepackage{amssymb} 
\usepackage{times}
\usepackage{graphicx}
\usepackage{graphicx}
\usepackage{dcolumn}
\usepackage{datetime}
\usepackage{soul}
\usepackage{subfigure}
\usepackage[bookmarks, colorlinks=true, plainpages = false,citecolor = red, urlcolor = black, 
filecolor = blue]{hyperref}
\usepackage{multirow} 
\usepackage[english]{babel}
\usepackage{graphicx}
\usepackage{amsmath}
 
\usepackage[english]{babel}
\usepackage{graphicx}
\usepackage{amsmath}
 
\begin{document}
\title{On the Nature of Tellurium Atomic Helices}
 
\author{George Kirczenow}  

\affiliation{Department of Physics, Simon Fraser
University, Burnaby, British Columbia, Canada V5A 1S6}

\date{\today}

\begin{abstract}\noindent

A theoreticsal study of single, double and triple hydrogen-terminated chains of tellurium atoms is presented.  Surprisingly, H-terminated single chains with 3 Te atoms per unit cell (as in Te crystals)  are found to be unstable. They relax to helices with lower energies and smaller twist angles. However, some compact disordered Te chains of mixed chirality are found to have still lower energies. Pairs of H-terminated Te atomic chains are found to form DNA-like double helices with lower energies than compact disordered structures of the two chains. Triplets of H-terminated Te atomic chains are found to form triple helices. The single, double and triple Te helices reported here are beyond the scope of previously studied periodic models with small unit cells.   
\end{abstract}

 \maketitle
 
 \section{Introduction}
\label{Intro}

Trigonal tellurium crystals consist of parallel helical chains of atoms. The unit cell contains three tellurium atoms. The interchain coupling between the tellurium atoms is weaker than the intrachain coupling. This suggests that nanostructures consisting of single helical chains of tellurium atoms and/or bundles of a few such chains may be possible.They would constitute unique examples of extreme helical nanowires exhibiting strong spin-orbit coupling. At present there is considerable interest in studying such tellurium nanostructures theoretically and synthesizing them.\cite{Bogomolov83MO,Bogomolov85MOZEO,Poborchii95MO,Inoue2005ZEO,GhoshDFT,GhoshDFTorigin,KobayashiCNT,MedeirosCNT,ChurchillExfol,HanRashba,IkemotoCNT,KramerDFT,QinCNTBNNT,YinDFTFET,SharmaDFT,HanDFT2022,PoborchiiZEOL,Paul2023}
Density functional theory (DFT) based studies of such nanostructures have been carried out. \cite{GhoshDFT,GhoshDFTorigin,MedeirosCNT,HanRashba,KramerDFT,QinCNTBNNT,YinDFTFET,SharmaDFT,HanDFT2022,Paul2023} 
Some examples have been realized experimentally, encapsulated in carbon nanotubes,\cite{KobayashiCNT,MedeirosCNT,IkemotoCNT,QinCNTBNNT} boron nitride nanotubes,\cite{QinCNTBNNT} zeolites\cite{Bogomolov85MOZEO,Inoue2005ZEO,PoborchiiZEOL} and mordenite.\cite{Bogomolov83MO,Bogomolov85MOZEO,Poborchii95MO,PoborchiiZEOL} Free-standing isolated tellurium atomic chains have not yet been synthesized although ultra-thin tellurium flakes have been made by exfoliation.\cite{ChurchillExfol} DFT-based studies of free standing\cite{GhoshDFT,HanRashba,KramerDFT,YinDFTFET,HanDFT2022,Paul2023} and encapsulated\cite{MedeirosCNT} tellurium atomic chains and few chain bundles have considered periodic infinite structures with small unit cells or few-atom (up to 12 Te atoms) chains, clusters or rings.\cite{GhoshDFTorigin,Pan2002} 

The present work complements the previous theoretical studies by calculating low energy geometries of free-standing {\em finite} single, double and triple tellurium atomic chains terminated with H atoms so as to passivate dangling bonds. Such systems with up to 49 Te atoms per atomic chain are considered. The low energy geometries of these systems are found with the help of DFT-based calculations. The low energy geometries found here differ qualitatively from those predicted if periodic infinite structures with small unit cells are assumed as in previous studies.  Surprisingly, it is found that single finite H-terminated Te chains with geometries similar to those of the chains in trigonal tellurium crystals are not stable and not even metastable -- they are unstable:  For long chains, they relax to regular helices with twist angles near $106^{\circ}$ per atom about the axis of the helix (except near the ends of the chain), as distinct from the $120^{\circ}$ twist angles in trigonal tellurium crystals. This novel finding is confirmed by DFT calculations presented here for infinite periodic Te chains that find that regular helices with 17 atoms per unit cell and twist angles near $106^{\circ}$ per atom have lower energies than do other regular helices. The present study also finds that some atomic chains with irregular unit cell geometries have marginally lower energies than those of regular helices. However, H-terminated Te atomic chains with compact disordered geometries are found to have still lower energies. These low energy compact disordered structures have mixed chiralities. Interestingly, in this work switching between local left and right chiralities during relaxation of the compact, disordered chains has been found to occur. It is also found here that pairs of H-terminated Te chains wrap around each other to form DNA-like double helix structures and that the energies of these double helices are lower than those of pairs of chains in compact disordered geometries. Triplets of H-terminated Te chains are found to relax to triple helices whose pitch is several times larger than that of the double helices, foreshadowing the crossover from Te nanowire to Te bulk crystal geometries. 

The remainder of this paper is organized as follows: In Section \ref{Method} the computational details are described. In Section \ref{Regular} it is shown that free standing H-terminated Te atomic chains with structures similar to those in Te crystals are unstable and relax to structures with smaller twist angles. In Section \ref{Periodic} regular infinite periodic Te atomic helices with various periods are studied. It is shown that geometries with {\em large} unit cells are the most stable of these structures and have twist angles consistent with with those of the relaxed H-terminated helices found in Section \ref{Regular}. In Section \ref{Irregular} infinite periodic and H-terminated irregular Te atomic chains are studied. Compact disordered Te atomic chains and their energetics are discussed in Section \ref{Disordered}. The chiralities of the Te atomic chains found in the present study are discussed in Section \ref{Chirality}. The geometries and energetics of double and triple Te atomic helices are presented in Section \ref{Double}. The present results are discussed further in Section \ref{Discussion} and some concluding remarks are presented in Section \ref{Conclusion}.
 
\section{Method}
\label{Method}

The low energy geometries of the systems considered here and their energies were found with the help of density functional theory (DFT) based calculations.
The DFT calculations were carried out with the Amsterdam Modeling Suite (AMS 2023) using the  Perdew-Burke-Ernzerhof generalized gradient approximation density functional, the ZORA/TZ2P basis and including relativistic effects at the level of spin-orbit coupling. The atomic geometries were optimized requiring the energy gradients to be 10$^{-4}$ Hartree/Bohr or less. The AMS 2023 FIRE optimization algorithm\cite{FIRE} was found to be suitable for the systems considered here.

\section{Results}
\label{Results}

\subsection{H-Terminated Single Atomic Chains}
\label{Regular}

\begin{figure}[t]
\centering
\includegraphics[width=0.7\linewidth]{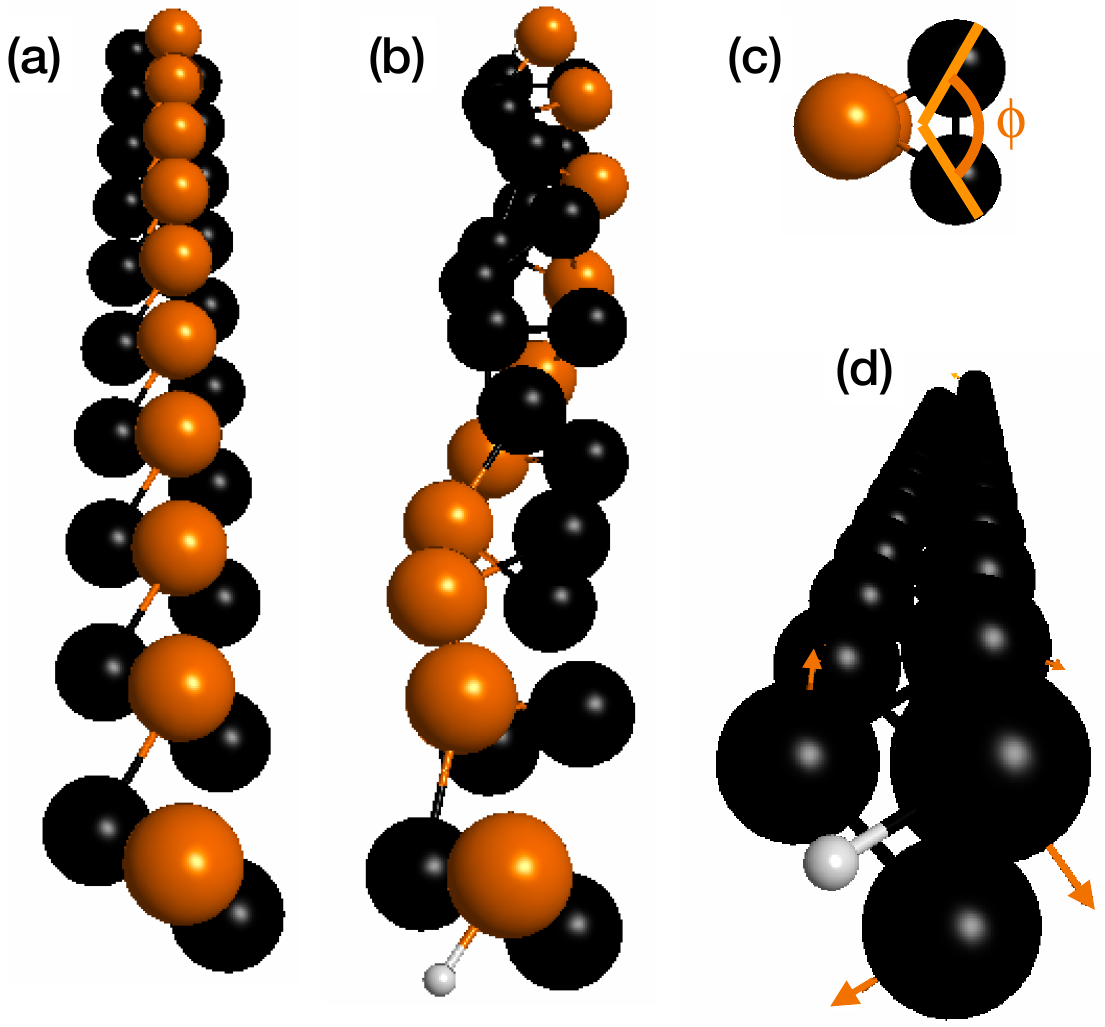}
\caption{
(a) 30 atom segment of infinite periodic Te atomic chain with 3 atoms per unit cell. Every third atom of the chain is colored orange in order to make the helical geometry obvious. The other Te atoms are black. (b) Relaxed Te$_{30}$H${_2}$ atomic chain. The dangling bonds at the ends of the Te chain are passivated with H atoms (gray). (c)The twist angle $\phi$ per atom about the helical axis of the atomic chain. The view is along the helical axis. (d) {\em Early} stage of the relaxation of a Te$_{48}$H${_2}$ atomic chain  starting from a geometry in which the Te atoms form a finite periodic structure with 3 atoms per unit cell similar to that in (a). The early displacements of the Te atoms are indicated by arrows whose lengths have been upscaled for clarity. Images prepared using Macmolplt software.\cite{MacMolPlt}
}
\label{chains} 
\end{figure}

Fig.~\ref{chains}(a) shows a segment of an isolated infinite periodic Te atomic chain with 3 atoms per unit cell optimized within DFT. The geometry is similar to that of a segment of a Te atomic chain in a trigonal tellurium crystal. In the present work the ends of such finite Te atomic chains were passivated by the addition of H atoms and the resulting structures were relaxed, optimizing their geometries and energies calculated within DFT. 
An example of such a relaxed structure (a relaxed Te$_{30}$H${_2}$ atomic chain) is shown in Fig.~\ref{chains}(b). In Fig.~\ref{chains}(a)-(c) every third atom is colored orange in order to help make the structure clear. 

Fig.~\ref{chains}(d) shows how the relaxation process begins for the case of a Te$_{48}$H${_2}$ atomic chain. The arrows (whose lengths are upscaled for clarity) show the displacements of the Te atoms from their initial positions early in the relaxation process. Fig.~\ref{chains}(d) shows that the {\em ends} of the finite periodic atomic chain with three atoms per unit cell are unstable against a torsional deformation. The deformation grows and propagates along the chain resulting finally in a regular Te$_{n}$H${_2}$ chain such as that in Fig.~\ref{chains}(b).  
This suggests that the results of calculations of relaxed Te atomic chain geometries that {\em impose} an infinite periodic chain geometry with a small unit cell may not be representative of physical atomic chains. This is because real atomic chains (other than closed rings) have ends, and the ends can exhibit local instabilities that propagate throughout the chain resulting in a qualitative global change in the chain's geometry.  More specifically, a calculation that assumes an infinite periodic chain with a 3-atom unit cell cannot detect such an instability since the instability nucleates at the ends of the chain whereas an infinite periodic chain has no ends.  

\begin{figure}[t]
\centering
\includegraphics[width=0.9\linewidth]{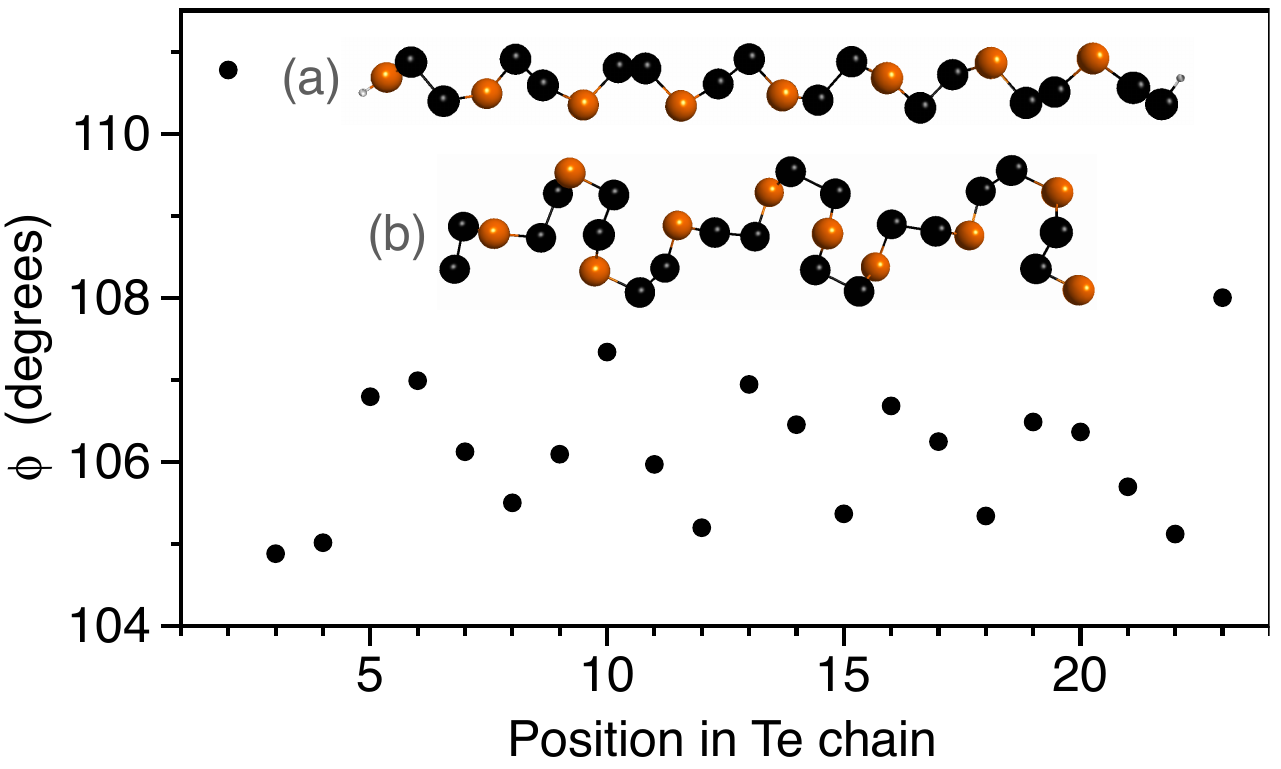}
\caption{
Black dots show the twist angle $\phi$ per atom (defined as in Fig.~\ref{chains}(c)) for a relaxed Te$_{24}$H${_2}$ regular helical atomic chain vs. position in the Te chain. Insets: (a) Relaxed regular Te$_{24}$H${_2}$ chain geometry. (b) 30 atom segment of a relaxed infinite irregular periodic Te atomic chain with 10 atoms per unit cell. Every third Te atom is colored orange for clarity.}
\label{twist} 
\end{figure}

Chain geometries such as that in Fig.~\ref{chains}(b) can be characterized by their twist angle $\phi$ per Te atom about the helical axis of the Te atomic chain. The meaning of the twist angle is made clear in Fig.~\ref{chains}(c) where two successive Te atoms of the chain are shown in black and the view is along the helical axis. For a periodic infinite Te helical chain with three atoms per unit cell as in Fig.~\ref{chains}(a), $\phi=120^{\circ}$. However, for relaxed Te$_{n}$H${_2}$ chains such as that in Fig.~\ref{chains}(b), $\phi$ is smaller. For the case of the Te$_{4}$H${_2}$ $\phi\sim114^{\circ}$. For longer regular chains $\phi$ decreases further as can be seen in Fig.~\ref{twist}. Here the twist angle $\phi$ for a relaxed Te$_{24}$H${_2}$ atomic chain vs. Te atom location in the chain is shown. (The chain itself is shown in inset (a) of Fig.~\ref{twist}.) Except near the ends of the Te chain, $\phi$ oscillates about an average value near 106$^{\circ}$ with an amplitude $\sim$$1^{\circ}$.

\begin{table}[b]
\caption{Calculated energies of some regular infinite periodic Te atomic helices with the same value of the atomic twist angle throughout the atomic chain. The number of atoms in the unit cell is $N$. $w$ is the winding number, i.e., the net number of complete 360 degree twists of the helix in one period of the atomic chain.  $\phi$ is the twist angle per atom  about the axis of the helix. $\epsilon$ is the energy per atom relative to that for the infinite periodic helix with three atoms per cell calculated within DFT.}
\begin{center}
\begin{tabular}{l|c|c|c|c}
 $N$ & $w$&  $\phi$& $\epsilon$ (meV) \\ 
 \hline
  $3$& $1$ & $120^{\circ}$ &0.0\\
  \hline
  $16$& $5$ & $112^{\circ}$ &-13.5\\
  \hline
   $13$& $4$ & $110.8^{\circ}$ &-15.1\\
  \hline
   $10$& $3$ & $108^{\circ}$ &-17.0\\
  \hline
    $17$& $5$ & $105.9^{\circ}$ &-17.4\\
  \hline
   $7$& $2$ & $102.9^{\circ}$ &-16.7\\
  \hline
 $14$& $4$ & $102.9^{\circ}$ &-16.7\\
  \hline
  $11$& $3$ & $98.2^{\circ}$ &-11.8\\
  \hline
 $15$& $4$ & $96^{\circ}$ &-7.9\\
  \hline
$4$& $1$ & $90^{\circ}$ &8.6\\
  \hline

\end{tabular}
\end{center}
\label{helices}
\end{table}

\subsection{Infinite Periodic Helices}
\label{Periodic}

For comparison, the calculated energies per atom for some regular infinite periodic Te atomic helices with various periods, winding numbers and twist angles $\phi$ per atom are presented in Table \ref{helices}.  The energy $\epsilon$ per atom is given relative to that for the infinite periodic helix with three atoms per cell. Notice that the structure with the lowest energy in Table \ref{helices} has a twist angle $\phi$ per atom close to $106^{\circ}$, consistent with the values of $\phi$ away from the ends of the relaxed Te$_{24}$H${_2}$ chain in  Fig.~\ref{twist}. This suggests that the helical structure in Fig.~\ref{chains}(b) forms because it is favored energetically. Note, however, that this lowest energy infinite periodic structure in Table \ref{helices} has a large unit cell of $N=17$ Te atoms and a winding number $w=5$. Te chains with such large periods have not been considered previously in the literature. 

Note also that in Table \ref{helices} the chain with the second smallest  number $N=4$ of Te atoms per unit cell has a higher energy $\epsilon$ per atom than the chain with $N=3$ atoms per unit cell. Thus studies restricted to infinite periodic chains with such small unit cells do not offer any hint that the $N=3$ atoms per unit cell structure may be unstable. By contrast, relaxation of a finite H-terminated $N=3$ atoms per unit cell chain
as in Fig.~\ref{chains}(d) proceeds by way of smoothly decreasing twist angles corresponding to decreasing energies, qualitatively consistent with the behavior of $\phi$ and $\epsilon$ as one scans downwards through the top half of Table \ref{helices}.

\subsection{Irregular Atomic Chains}
\label{Irregular}

The structures listed in Table \ref{helices} are regular in the sense that each such structure can be described by a single value of the twist angle $\phi$. Low energy {\em ir}regular infinite periodic Te atomic chains that do not have this simple property are also possible. The lowest energy relaxed irregular infinite periodic structure found in the present study has a unit cell of $N=10$ Te atoms and an energy of -19.7 meV per atom, relative  to that for the infinite periodic helix with three atoms per cell. This energy is marginally lower than that of the lowest energy regular periodic Te helix in Table \ref{helices}. A 30 atom segment of this infinite irregular periodic Te atomic chain is shown in inset (b) of Fig.~\ref{twist}. 

\begin{figure}[t]
\centering
\includegraphics[width=0.99\linewidth]{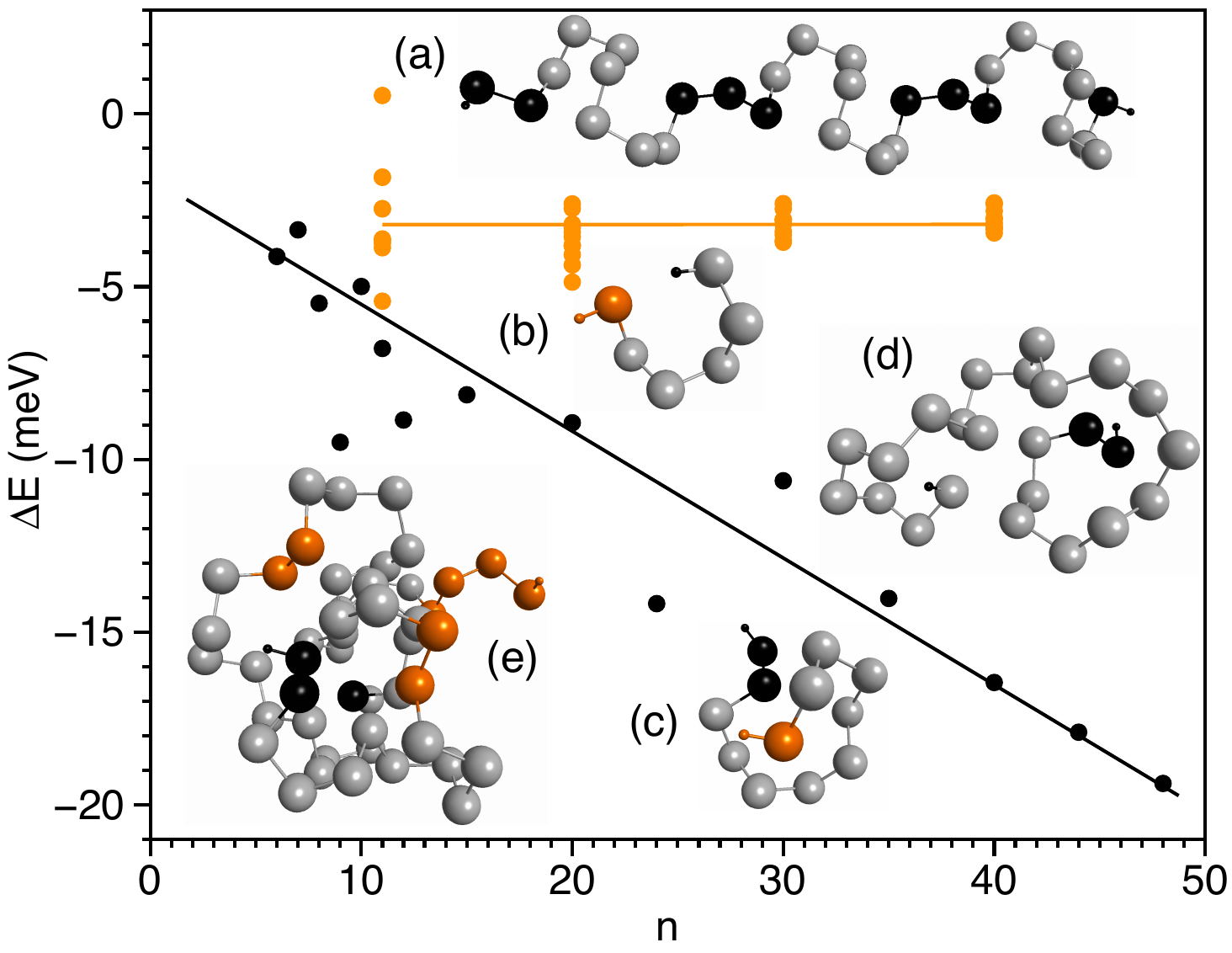}
\caption{
Energy difference per Te atom $\Delta E$ between irregular and regular Te$_{n}$H${_2}$ chains (orange dots) and between compact disordered and regular Te$_{n}$H${_2}$ chains (black dots) vs. number of Te atoms $n$. Straight lines are guides to the eye. Inset (a): Relaxed geometry of irregular Te$_{30}$H${_2}$ chain. Insets  (b), (c), (d) and (e) show the relaxed geometries of examples of compact disordered Te$_{n}$H${_2}$ chains for $n=6, 12, 24~\text{and}~48$, respectively. Atoms colored black, orange and gray indicate locations of right handed, left handed and mixed chirality, respectively.}
\label{compact} 
\end{figure}

Segments of various lengths $n$ of this infinite irregular periodic Te atomic chain were terminated with H atoms, and then relaxed and their energies $E_\text{irreg}$ were calculated within DFT. An example of such an H-terminated irregular structure with $n=30$ Te atoms is shown in inset (a) of Fig.\ref{compact}. The energies of these irregular H-terminated structures were compared with the calculated energies $E_\text{reg}$ of relaxed regular Te$_{n}$H${_2}$ chains with the same numbers $n$ of Te atoms. The differences in energy per Te atom $\Delta E = (E_\text{irreg}-E_\text{reg})/n$ between the irregular and regular Te$_{n}$H${_2}$ chains are plotted in orange vs. the number $n$ of Te atoms in the chain in Fig.\ref{compact}. The orange straight line is a guide to the eye. The energies per Te atom of the H-terminated irregular structures are lower on average than those of the regular structures by $\sim 3$meV, consistent with the difference of $\sim 2.2$meV for the corresponding infinite periodic structures; the spread in values of $\Delta E$ for each fixed number $n$ of Te atoms in the chain in  Fig.\ref{compact} is due to different locations of the ends of the irregular chains in the unit cells of the parent periodic structure.

\subsection{Compact Disordered Chains}
\label{Disordered}

The irregular Te$_{n}$H${_2}$ chains discussed above have marginally lower energies than their regular counterparts. However the present study reveals that some compact disordered Te$_{n}$H${_2}$ chains have still lower energies, $E_\text{c.d.}$. Examples of relaxed compact disordered Te$_{n}$H${_2}$ chains for $n=6, 12, 24~\text{and}~48$ are shown in insets (b), (c), (d) and (e) of  Fig.\ref{compact}, respectively. The differences in energy per Te atom $\Delta E = (E_\text{c.d.}-E_\text{reg})/n$ between some compact disordered and regular Te$_{n}$H${_2}$ chains are plotted in black vs. the number $n$ of Te atoms in the chain in Fig.\ref{compact}. The black straight line is a guide to the eye. Many metastable compact disordered Te$_{n}$H${_2}$ chain geometries are possible for each value of $n$ and it is not feasible to identify all of them. In the present study a few compact disordered geometries were sampled and relaxed for each value of $n$ included in Fig.\ref{compact}, and the energy values shown are the lowest ones found. Thus the energy values plotted should be viewed as an upper bound on the lowest energy for each $n$. However, the results presented are sufficient to demonstrate that for each $n$ there exist compact disordered structures whose energies are lower than those of both the regular and irregular structures considered here.

The difference in energy per Te atom $\Delta E = (E_\text{c.d.}-E_\text{reg})/n$ between the compact disordered and regular Te$_{n}$H${_2}$ chains is seen in Fig.\ref{compact} to scale roughly linearly with $n$. This can be attributed physically to the attractive interaction of the compact disordered chain with itself due to different parts of the chain being in close proximity. At the level of the simplest mean field theory, this contribution to the chain's self-interaction energy would be proportional to $n^2$ and hence the corresponding energy per Te atom would be proportional to $n$, consistent with the roughly linear behavior of $\Delta E = (E_\text{c.d.}-E_\text{reg})/n$ vs. $n$ in Fig.\ref{compact}. This is also consistent with the qualitative finding of the present study (not shown in Fig.\ref{compact}) that Te$_{n}$H${_2}$ chains with less compact relaxed geometries tend to have higher energies than those with more compact geometries for the same value of $n$. 

\subsection{Chirality}
\label{Chirality}

Regular Te$_{n}$H${_2}$ chains are chiral, i.e., they are right handed or left handed throughout the chain. As will be seen below this is not the case for the irregular chains or low energy compact disordered chains considered here. To decide whether an atomic chain is right or left handed at a particular location it is necessary to consider at least 4 atoms around that location since any group of 3 or fewer atoms has mirror symmetry. However the 4 consecutive atoms that include a particular atom of a chain can be chosen in 4 different ways. These 4 chosen sets of atoms may be of the same handedness or of mixed handedness, i.e., some sets may be right handed while others are left handed. This is exemplified in the insets of Fig.\ref{compact} where atoms at purely right (left) handed locations are shown in black (orange) and those at mixed chirality locations are shown in gray. There are more mixed chirality sites than pure right or left chirality sites in all of the cases shown in  insets of Fig.\ref{compact}. Interestingly, the local handedness is not necessarily preserved when a compact disordered structure is relaxed to optimize its geometry and energy calculated within DFT. Even if the starting disordered structure is fully right handed, left handed local defects can appear in it during relaxation. 

\subsection{Double and Triple Helices}
\label{Double}

\begin{figure}[t]
\centering
\includegraphics[width=0.99\linewidth]{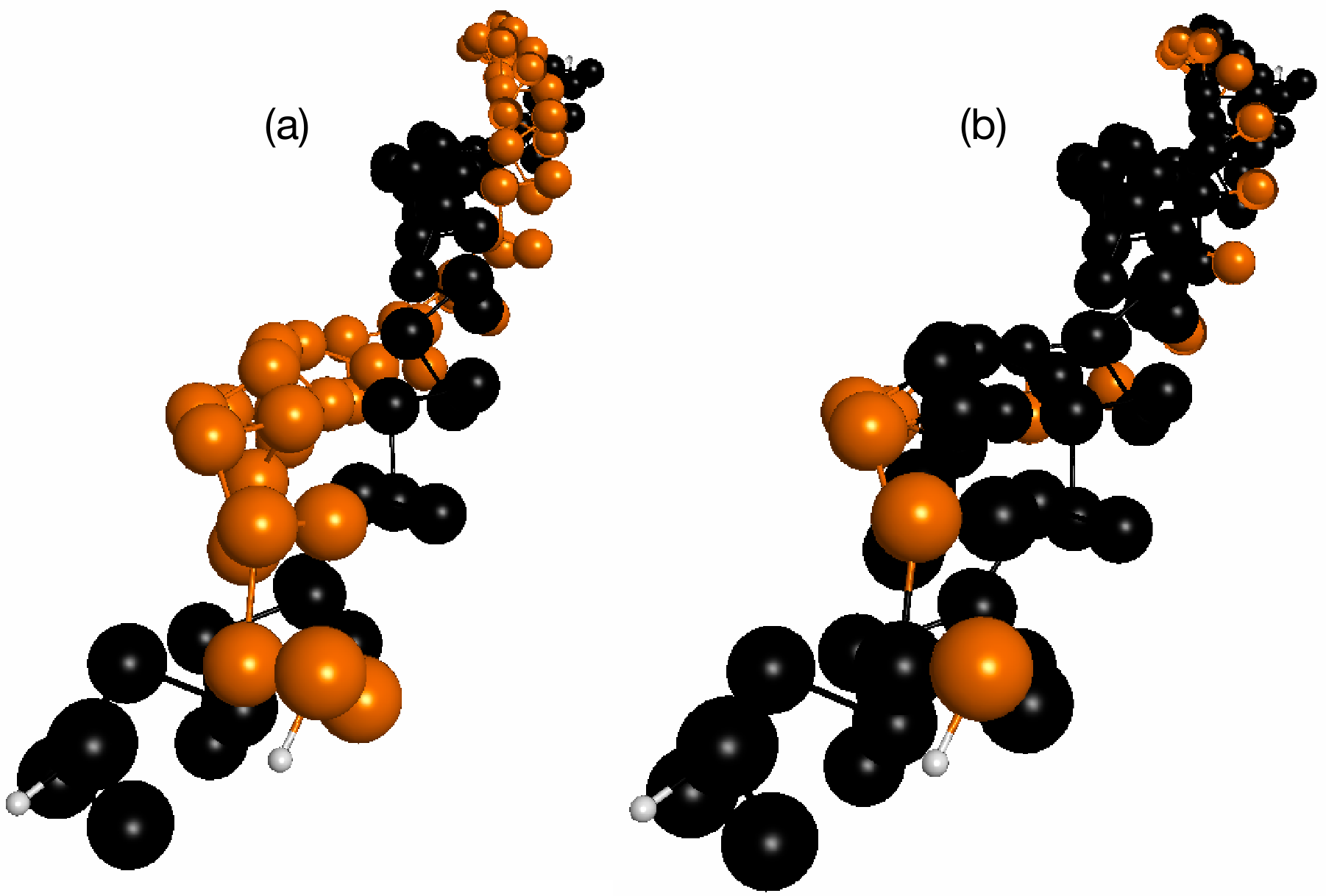}
\caption{
Relaxed geometry of a Te$_{98}$H${_4}$ double helix consisting of two Te$_{49}$H${_2}$ atomic chains. The colors help understand the structure.  (a) Te atoms of one chain are orange. Those of the other chain are black. (b) Every third Te atom of one chain is orange. The other Te atoms are all black.}
\label{dbl} 
\end{figure}

The present work also investigated the possible structures of pairs and triplets of parallel  Te$_{n}$H${_2}$ atomic chains. Upon optimizing their energies within DFT, these were found to relax to double and triple helices respectively. An example (a Te$_{98}$H${_4}$ double helix) is shown in Fig.\ref{dbl}. Here two Te$_{49}$H${_2}$ atomic chains wrap around each other in a DNA-like fashion. A single complete turn of the double helix is comprised of approximately 74 Te atoms. 

In the present study the calculated energies of some Te$_{2n}$H${_4}$ double helices were compared with those of corresponding pairs of Te$_{n}$H${_2}$ chains with compact disordered geometries. The double helices were found to have lower energies than the compact disordered structures. In Sec. \ref{Disordered} the finding that, for single Te$_{n}$H${_2}$ atomic chains, compact disordered chain geometries have lower energies than regular helices was attributed to attractive interactions between different parts of the compact disordered chain that are close to each other. This explanation is consistent with the present finding that {\em double} helices have lower energies than the corresponding compact disordered structures. This is because in the double helix the two chains are in close proximity with each other along the entire length of each chain whereas in compact disordered structures close proximity between the two chains or between different parts of the same chain is not as common. This suggests that experimental realization of these double helices may be less challenging than that of single chain regular helices.
Interestingly, experimental realization of double Te helices encapsulated in carbon nanotubes has been reported.\cite{KobayashiCNT} However, the double helix structures proposed by Kobayashi and Yasuda based on their imaging data\cite{KobayashiCNT} were simpler than that shown in Fig.\ref{dbl}.  

\begin{figure}[t]
\centering
\includegraphics[width=0.99\linewidth]{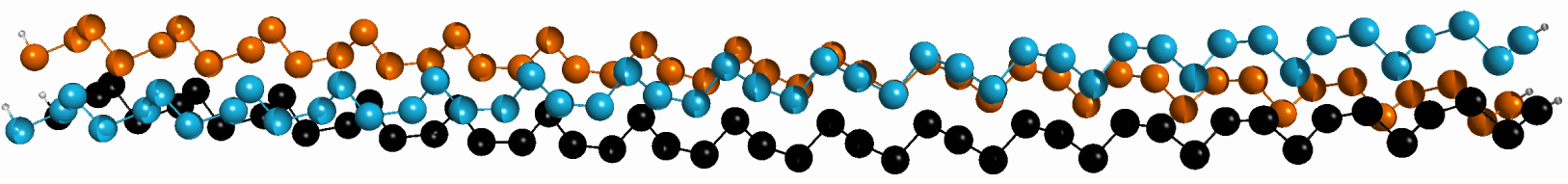}
\caption{
Relaxed geometry of a Te$_{144}$H${_6}$ triple helix consisting of three Te$_{48}$H${_2}$ atomic chains. The Te atoms of the three chains are colored blue, orange and black, respectively, for clarity. }
\label{trpl} 
\end{figure}

The relaxed geometry of a Te$_{144}$H${_6}$ triple helix is shown in Fig.\ref{trpl}. The Te atoms of the three Te$_{48}$H${_2}$ atomic chains making up the helix are colored blue, orange and black, respectively. Inspection of the left and right hand ends of the triple helix in Fig.\ref{trpl} shows the blue (orange) chain at the top on the right (left).  It follows that the 144 Te atoms shown in Fig.\ref{trpl} (48 Te atoms per chain) comprise roughly 1/3 of a complete turn of the triple helix about its axis. 

This should be compared with the double helix in Fig.\ref{dbl} where approximately 74 Te atoms (37 Te atoms per chain) comprise a complete turn of the double helix about its axis. Thus the pitch of the triple helix is several times larger than that of the double helix. The twist angles $\phi$ of the individual atomic chains of the triple helix are $\sim 118^{\circ}$, approaching the bulk value $120^{\circ}$ of trigonal tellurium crystals. This reveals how the geometry of the Te nanowire approaches that of a macroscopic Te crystal as the number of Te atomic chains making up the nanowire increases.

The present study of tellurium nanowires has also yielded a significant insight into the basic physics of crystalline tellurium itself. In particular, the present study has shown that the simple intrachain structure (3 atoms per unit cell) of the atomic chains that make up crystalline tellurium is stabilized by the relatively weak interchain interactions between atoms belonging to different chains: If the interchain interactions were turned off the present study shows that the intrachain structure would switch to approximating regular helices with 17 atoms atoms per cell or to irregular helices or become disordered. 

\section{Discussion}
\label{Discussion}

The density functional theory based study presented here has explored the structures and energetics of nanostructures consisting of single, double and triple hydrogen-terminated chains of tellurium atoms. The single chains studied were found to exhibit regular or irregular helical structures, or disordered structures. For long atomic chains, away from their ends the regular helical structures were found to exhibit twist angles close to those of infinite periodic Te helices with 17 atoms per unit cell. The H-terminated regular and  irregular helical structures were found to be metastable, whereas H-terminated helical chains with structures similar to the chains with 3 atoms per unit cell that make up trigonal tellurium crystals were found to be unstable, the instability propagating through the chain beginning at its ends. Compact disordered H-terminated chains with mixed chiralities were found to have lower energies than either the regular or irregular H-terminated helical structures due to attractive interactions between different parts of the disordered chain. Pairs of H-terminated Te atomic chains were found to form DNA-like double helices with the two chains wrapped around each other. The double helices were found to have lower energies than compact disordered structures of the two chains due to attractive interactions between the two chains of the double helix. Triplets of H-terminated Te atomic chains were found to form triple helices whose pitch is several times larger than that of the double helix. These findings have not been anticipated by previous DFT based calculations that were limited to studies of simpler geometries, i.e., periodic infinite structures with small unit cells or small numbers of atoms. These restrictions precluded consideration of the complex single chain geometries as well as of the double and triple helices proposed here. 

\section{Conclusion}
\label{Conclusion}

The present work suggests that the low energy geometries of tellurium atomic nanowires are much more complex and interesting than previously thought. Experiments testing the present predictions would be highly desirable. To date there have been no experimental studies of free-standing tellurium atomic chains or few chain bundles. Perhaps the present work will provide encouragement and some guidance for such experiments.   
A possible, as yet untried, approach to experimental synthesis of tellurium nanowires may be to employ passivation of Te dangling bonds by hydrogen atoms (as in the present theoretical study) or by other species. Alternatively,  
isolated Te nanowires may be regarded as a limiting case of nanowires encapsulated in carbon nanotubes with large diameters. Thus some of the nanowire geometries predicted by the present work may possibly occur in some of those experimentally realized systems.  Analysis of experimental data in the light of the present theoretical results may therefore be interesting.

\begin{acknowledgments}
This research was supported by the Digital Research Alliance of Canada.
\end{acknowledgments}
%% ----------------------------------------------------------------------------------------------------
{

\end{document}